\documentclass[%showpacs, 
twocolumn, preprintnumbers, amsmath,amssymb,nofootinbib, superscriptaddress]{revtex4}
\usepackage{graphicx}% Include figure files
\usepackage{dcolumn}% Align table columns on decimal point
\usepackage{bm}% bold math
\usepackage{footmisc}
\newcommand{\bra}{\langle}
\newcommand{\ket}{\rangle}

\newcommand\ee{\end{equation}}
\newcommand\be{\begin{equation}}
\newcommand\eea{\end{eqnarray}}
\newcommand\bea{\begin{eqnarray}}
\newcommand{\sfrac}[2]{{\textstyle\frac{#1}{#2}}}

\begin{document}

%\preprint{}

\title{A relativistic non-relativistic Goldstone theorem:\\
gapped Goldstones at finite charge density}% Force line breaks with \\

\author{Alberto Nicolis}
%\email{nicolis@phys.columbia.edu}
%
\affiliation{%
Physics Department and Institute for Strings, Cosmology, and Astroparticle Physics,\\
Columbia University, New York, NY 10027, USA
}%

\author{Federico Piazza}
%\email{fpiazza@apc.univ-paris7.fr}
%
\affiliation{%
 Paris Center for Cosmological Physics and Laboratoire APC,\\  Universit\'e Paris 7, 75205 Paris, France}%


\begin{abstract}
We adapt the Goldstone theorem to study spontaneous symmetry breaking in relativistic theories at finite charge density. It is customary to treat systems at finite density via non-relativistic Hamiltonians. Here we highlight the importance of the underlying relativistic dynamics. This leads to seemingly new results whenever the charge in question is spontaneously broken \emph{and} does not commute with other broken charges. We find that that the latter interpolate \emph{gapped}  excitations. 
In contrast, all existing versions of the Goldstone theorem predict the existence of gapless modes.
We derive exact non-perturbative expressions for their gaps, in terms of the chemical potential and of the symmetry algebra.
\end{abstract}

%\pacs{\dots}% PACS, the Physics and Astronomy
\maketitle

%%%%%%%%%%%%%%%%%%%%%%%%%%%%%%%%%%%%%%%%%%%%%%%%%%%%%%%%%
%%%%%%%%%%%%%%%%%%%%%%%%%%%%%%%%%%%%%%%%%%%%%%%%%%%%%%%%%

%\section{Preliminary considerations}
\noindent
{\em Preliminary considerations.}
Non-relativistic Goldstone theorems~\cite{Lange, NC,brauner,kaon2,kaon,brauner2} display interesting twists compared to the standard relativistic one~\cite{goldstone,weinberg}. First, the non-relativistic version is notoriously less powerful. For instance, it guarantees the existence of gapless {\em zero-momentum} excitations, but says nothing about their properties, like e.g.~stability, at finite momenta. A physically relevant example is phonons in superfluid helium-4, which are unstable with a decay rate $\Gamma \sim k^5$---that is, they are not really eigenstates of the Hamiltonian. Second, as far as counting is concerned,
a classic result by Nielsen and Chadha \cite{NC} states that for non-relativistic systems, the number of gapless Goldstone excitations equals the number of spontaneously broken generators, provided one counts the Goldstone excitations with {\em even} dispersion law (e.g.~$\omega\sim k^2 $) {\em twice}. 

However, to the best of our knowledge, all fundamental interactions are described by relativistic field equations. This means that for the so-called non-relativistic systems in the real world, Lorentz invariance is broken only {\em spontaneously}, i.e.~by the state of the system, rather than at the level of the dynamics. It is thus tempting to ask whether the very constrained framework of relativistic field theories can give non-trivial insights into physical systems that are effectively non-relativistic---and in particular, whether it can be used to sharpen or correct the non-relativistic versions of the Goldstone theorem. 

An immediate reaction to this idea is that in no way can relativistic effects be relevant for systems that, like condensed matter systems in the lab, have a very non-relativistic equation of state, a very small speed of sound compared to that of light, and so on. But there is more to relativity than just the so-called ``relativistic effects'', which are weighed by $(v/c)^2$. First, there is the statement of relativity itself---that all inertial frames are equivalent---which is valid, and powerful, even in the $c \to \infty$ limit, corresponding formally to the Galilean limit of Lorentz invariance.
Then, there are properties of relativistic field theories that are not directly statements of symmetry, but that are nevertheless crucial for the consistency of the theory. 
For instance, we will see below that the vanishing of commutators for space-like separated local operators can be used to remove an assumption of the Nielsen-Chadha theorem. Finally---and this will be the useful aspect for our purposes---the fundamental relativistic viewpoint offers an unambiguous starting point to analyze the pattern of spontaneous symmetry breaking, for spacetime symmetries and internal ones.

%An immediate reaction to this idea is that the spontaneous breakdown of Lorentz invariance also entails the existence of Goldstone bosons. For solids, these can be identified with the phonons. For fluids, with the sound waves and the shear waves (or vortices \cite{ENRW}).
%The interactions of these Goldstones---among themselves and with the other degrees of freedom---are strongly constrained by the spontaneously broken Lorentz symmetry. But apart from this, if one focuses on other degrees of freedom and is not interested in the dynamics of these Goldstones, then ``anything goes'', in the sense that Lorentz invariance has no further implications.

To clarify this last statement with an example, let's consider directly the system we want to focus on for the rest of paper: a relativistic theory with Hamiltonian $H$ and a group of internal symmetries, at finite density for one of the corresponding charges, $Q$.
%The non-relativistic systems that we consider in this paper are particularly transparent to the underlying full-relativistic description and allow to clarify the above point of view with a concrete example. We consider a relativistic theory with a group of internal symmetries, at finite density for one of the corresponding conserved charges, $Q$. 
The ground state $| \mu \ket$ (i.e. the state of minimal energy for given average charge density) can be found by the method of Lagrange multipliers as the state minimizing the modified Hamiltonian
%\be \label{hamiltonian}
$\tilde H = H - \mu Q$,
%\ee
where $\mu$ is the chemical potential\footnote{In general, the r.h.s.~should read $\lambda | \mu \ket$, with non-vanishing $\lambda$. From applying the familiar thermodynamic relation $E+ PV = T S + \mu Q$ to our zero-temperature system,  we see that $\lambda$ is related to the pressure. Such a charge density breaks Lorentz invariance, because it transforms like the time component of a four-vector. 
In the following we will drop $\lambda$ from our formulae, to make them less cluttered, although keeping track of it is straightforward. Formally,
for any given $\mu$, we can set $\lambda$ to zero by adjusting the cosmological constant.},
%, which changes $E$ and $P$ individually while preserving $E+PV$.}
\be \label{ssp}
\tilde H | \mu \ket = \big( H - \mu Q \big) |\mu \ket = 0 \; .
\ee
%The conclusion is that all our results below apply once we shift the spectrum of $\tilde H$ by $\lambda$. Equivalently, for any given $\mu$, we can set $\lambda$ to zero by adjusting the cosmological constant, which changes $E$ and $P$ individually while preserving $E+PV$.
It is standard practice to use the non-relativistic Hamiltonian $\tilde H$ to study the system at finite density~\cite{lebellac}. However, it is clear from the outset that we are in the presence of a \emph{spontaneous}---rather than explicit---breaking of Lorentz symmetry. Introducing $\tilde H$ is purely a mathematical tool to find a state with the desired properties. The Hamiltonian of the system is still $H$. Heisenberg picture operators evolve in time as dictated by $H$. 
In order to better understand the role of $\tilde H$, consider the case in which \emph{also} the (internal) symmetry generated by $Q$ is spontaneously broken, \emph{i.e.} 
\be \label{breaking}
\bra \mu | [Q,A(x)] | \mu \ket \neq 0 
\ee
for some order parameter $A(x)$. Then $| \mu \ket$ cannot be an eigenstate of $Q$, because this would be 
inconsistent with~\eqref{breaking}.
But since $| \mu \ket$ obeys eq.~\eqref{ssp}, it cannot be an eigenstate of $H$ either.  
We conclude that, at finite density for $Q$, if $Q$ is spontaneously broken, so is $H$ \cite{NP}. 
This means that if $Q$ is broken, we cannot classify the states of the system---including our ground state $| \mu \ket$---as eigenstates of the fundamental Hamiltonian $H$. The best we can do is trying to diagonalize the unbroken combination $\tilde H = H - \mu Q$. $| \mu \ket$ is the eigenstate with lowest eigenvalue.  The excitations of the system---including the Goldstone bosons---will correspond to higher eigenstates.

%It is standard practice to use the non-relativistic Hamiltonian $\tilde H$ to study the system at finite density~\cite{lebellac}. 
%However, it is clear from the outset that we are in the presence of a \emph{spontaneous}---rather than explicit---breaking of Lorentz symmetry. 
%This is particularly apparent in the case when the symmetry generated by $Q$ is broken by the state $|\mu\ket$, that is, when there exists an order parameter $A$ such that
%\be \label{breaking}
%\bra \mu | [Q,A] | \mu \ket \neq 0 \, .
%\ee
%In fact, the above is inconsistent with $Q | \mu \ket = q | \mu \ket$, with real $q$.
%But since $| \mu \ket$ obeys eq.~\eqref{ssp}, it cannot be an eigenstate of $H$ either.  
%That is, at finite density for $Q$, if $Q$ is spontaneously broken, so is $H$ \cite{NP}. 
%This means that if $Q$ is broken, we cannot classify the states of the system---including our ground state $| \mu \ket$---as eigenstates of the fundamental Hamiltonian $H$. The best we can do is try to diagonalize the unbroken combination $\tilde H = H - \mu Q$. $| \mu \ket$ is the eigenstate with lowest eigenvalue (zero, after we drop $\lambda$).  The excitations of the system---including the Goldstone bosons---will correspond to higher eigenstates.

Since the Nielsen-Chadha theorem implicitly assumes that Heisenberg picture operators evolve in time with the same (non-relativistic) Hamiltonian that is minimized by the ground state, we conclude that that theorem {\em does not apply} to systems that are `non-relativistic' because of a finite density for a charge that is spontaneously broken.
Indeed, for such systems it is not even clear how one would study the spontaneous breaking of $Q$ in a setup that is non-relativistic from the start. There are systems of this sort where $Q$ is broken `before' (i.e., at higher energy scales) Lorentz invariance is: any relativistic theory with ordinary spontaneous symmetry breaking in its Poincar\'e-invariant vacuum, can be put in a state of arbitrarily low density for the broken charge \cite{NP}.

When there are local operators obeying~\eqref{breaking}, an alternative viewpoint suggests itself. By eq.~\eqref{ssp}, the action of the Hamiltonian on $|\mu\ket$ is proportional to that of the symmetry generator. We are thus in the presence of a state that evolves in time along a symmetry direction, at `speed' $\mu$.
Any field  $\phi_j(x)$ that transforms non-trivially under the symmetry  can then feature a spacially homogeneous, time-dependent expectation value, obeying
\be
\frac{d}{dt}\langle {\dot \phi}_j \rangle = \mu \langle \delta \phi_j \rangle \; ,
\ee
where $\phi_j \rightarrow \phi_j + \delta \phi_j$ is the action of the symmetry in field space. 
%
% More formally, from \eqref{ssp} one easily derives  that any local operator has an expectation value on $|\mu\ket$ that either is constant (when it is `unbroken') or evolves in time proportionally to the action of the symmetry generator $Q$. As a consequence, the state $|\mu\ket$ corresponds, at the semi-classical level, to spacially homogeneous and time dependent  field configuration  $\phi_j = \phi_j(t)$ satisfying
%\begin{equation}
%{\dot \phi}_j(t) = \mu \, \delta \phi_j,
%\end{equation}
%$\phi_j\rightarrow \phi_j + \delta \phi_j$ being the action of the symmetry in field space. 
In~\cite{NP}, we dubbed this situation `spontaneous symmetry probing' (SSP)---there, we were using `$c$' in place of `$\mu$'.
This viewpoint is particularly useful in the semiclassical limit, where we can think of time evolution in terms of classical trajectories in field space.
We refer the reader to \cite{NP} for details.

\vspace{.3cm}

%\section{Assumptions and formalism}
\noindent
{\em Assumptions and formalism.}
We now want to study the low-energy spectrum of the system at finite density, by considering the Goldstone states associated with $Q$ and with other broken generators. Let's assume that the theory enjoys a Lie 
group of internal symmetries, with generators $Q_1, Q_2, \dots , Q_N$. Without loss of generality we can set the first generator to be our $Q$, $Q=Q_1$. The remarks we  made above lead to the following  hypotheses:
%\begin{enumerate}
%\item 
{\em (a)} The Heisenberg-picture currents evolve in time as dictated by the original Hamiltonian, i.e.
%\begin{equation} \label{ass2}
$J^\mu_a(t, {\vec x}) = e^{i (H t - \vec P \cdot \vec x)} J^\mu_a(0) e^{-i (H t - \vec P \cdot \vec x)}$,
%\end{equation}
where $a = 1,\dots, N$, and $\vec P$ is the total momentum operator.
%\item 
{\em (b)} The state $|\mu \ket$ is the ground state of $\tilde H = H - \mu\, Q$. 
%\end{enumerate}
The crucially different role played by $H$ and $\tilde H$ is the origin of the discrepancy between our results and the existing literature on non-relativistic Goldstone theorems, e.g.,~\cite{NC,brauner,kaon2,kaon}.
   
Consider then the case in which the first $n$ of the $Q_a$'s---including $Q_1$---are spontaneously broken. By definition, for each spontaneously broken $Q_a$, there must exist a local operator $ A_I (x)$---an `order parameter'---that makes the matrix element
\be
\kappa_{aI} \equiv \langle \mu |[Q_a(t), A_I (0)] | \mu \rangle 
\ee
nonzero. The index $I=1, \dots, m \le n$ in general runs over fewer values than the number of broken generators, simply because the same $A_I(x)$ typically serves as an order parameter for two or more symmetries. 

In the matrix element above, $A_I(x)$ is evaluated at the (space-time) origin. From now on, to simplify the notation, whenever a local operator is evaluated at the origin, we will drop its argument, ${\cal O}(0) \to {\cal O}$. $Q_a$ is formally evaluated---in Heisenberg picture---at time $t$. But since $Q_a$ commutes with the Hamiltonian, it is constant in time, and so is the $\kappa_{aI}$ matrix element. 
To convince oneself that this is true even though spontaneously broken charges are not completely well-defined operators, one can use the local conservation of the current:
\be
 \int \! d^3 x \langle \mu| [\dot J_a^0 (\vec{x},t) , A_I] |\mu\rangle + \! \int \!d^3 x \langle \mu| [\partial_i J_a^i (\vec{x},t) , A_I] |\mu\rangle = 0 \; .\nonumber
\ee
%\begin{multline}
% 0 = \!  \int \! d^3 x \langle \mu| [\partial_\mu J_a^\mu (\vec{x},t) , A_I] |\mu\rangle \\
% =  \! \int \! d^3 x \langle \mu| [\dot J_a^0 (\vec{x},t) , A_I] |\mu\rangle + \! \int \!d^3 x \langle \mu| [\partial_i J_a^i (\vec{x},t) , A_I] |\mu\rangle \; .\nonumber
% \end{multline}
The first term is the time-derivative of our $\kappa_{aI}$, while the second is a boundary term that only receives contributions from spacial infinity. Since for relativistic QFTs the commutator of space-like separated local operators vanishes---and we are breaking Lorentz symmetry only spontaneously---such a term is guaranteed to vanish\footnote{This is the removal of one of the Nielsen-Chadha assmputions we alluded to above.}.
We conclude that $\kappa_{aI}$  is constant in time.

We now use assumptions {\em (a)} and {\em (b)} above to `pull out' of $J^0_a$ its $\vec x$-  and $t$-dependence:
\begin{eqnarray} \label{step2}
\kappa_{aI} \!&=& \! \int \! d^3 x \langle \mu| J_a^0 (\vec{x},t) A_I |\mu\rangle  - {\rm c.c.}  \\
\!&=& \! \int \! d^3 x \langle \mu| e^{i  \mu Q  t} \, J_a^0 \, e^{-i (H t - P\cdot \vec{x})} A_I |\mu\rangle  - {\rm c.c.}\nonumber\, ,
\end{eqnarray} 
where we used that spacial translations are {\em not} spontaneously broken, $\vec P | \mu \ket = 0$, and we are assuming that both $J^0_a$ and $A_I$ are hermitian operators.
%To simplify the notation, from now on we will drop the superscript `$0$' from the time component of the current, as well as its argument $(0)$.
Inside the matrix element, we now insert a complete set of intermediate momentum eigenstates $|n,\vec{p} \, \rangle$, 
where $n$ labels other quantities that characterize these states. Schematically, 
\begin{eqnarray} \label{completeness}
& & \langle \mu | e^{i  \mu Q  t} \, J^0_a \, e^{-i (H t-P\cdot \vec{x} )} A_I | \mu \rangle   \\
& = & \sum_{n,p} e^{i \vec{p} \cdot \vec{x}} \langle \mu | e^{i  \mu Q  t} J^0_a e^{-i \mu Q  t} e^{- i \tilde H t}  |n,\vec{p} \, \rangle \langle n,\vec{p} \, | \, A_I |\mu \rangle \nonumber
\end{eqnarray}
where we  rewrote $H$ in terms of $\tilde H$ and $Q$.
%, and we assumed a non-relativistic normalization for the states, $\bra n, \vec p \,  | n', \vec p \, ' \ket = \delta_{nn'} \delta_{\vec p \vec p \, '}$. 
The Hamiltonian $\tilde H$ commutes with the spatial momentum, because $H$ and $Q$ do. We can then choose the $|n, \vec p \ket $ states to be eigenstates of $\tilde H$ as well, with eigenvalues $E_n(\vec p \, )$. 
The integral in $d^3 x $ projects onto the zero momentum states, and we are left with
\be
\kappa_{aI} = \sum_n  e^{- i E_n(0) t} \langle \mu | e^{i \mu Q t} J^0_a e^{-i \mu Q t}   |n,0 \rangle \langle n,0  |  A_I |\mu \rangle - {\rm c.c.} \label{step3}
\ee 
%Assuming that the commutators of charges and currents are simply
%%%
%%%
%\footnote{Since the analog of \eqref{structure} has to hold at the level of the charges, $[Q_a,Q_b] = i f^c_{ab} Q_c$,
%the only possible additions to \eqref{structure} are terms that vanish once we integrate in $d^3 x$---that is, total spacial derivatives. However, since we did integrate in $d^3 x$, or equivalently, we took the $\vec p \to 0$ limit, such possible extra terms will not matter for us either.}
%%%
%%%
%\be \label{structure}
%[Q_a, J^0_b(x)] = i f_{ab}^c \, J_c^0 (x) \; ,
%\ee

The constancy in time of the above expression gives important information about the spectrum of the theory in the limit of zero spatial momentum. There are two distinct cases to consider, depending on whether  $Q_a$ commutes with $Q \equiv Q_1$ or  not.  
For brevity, let us call these two classes of generators `C' and `NC', short for `commuting' and `non-commuting'.
 We will show that C-generators interpolate states $|n,\vec{p} \, \rangle$ that have zero energy in the limit of zero momentum, $E_n(0) = 0$. %Such Goldstone states are, as maybe expected, gapless but not necessarily ``massless" in the Lorentz invariant sense, as they can display Lorentz breaking dispersion relations~\cite{NP}. 
Vice versa, %If $J_a$ does not commute with $J_1$ then the situation is slightly more complicated. We will show that we can find a basis in the space of the NC generators such that~\eqref{step3} can conveniently be expressed through a time dependent combination of \emph{pairs} of currents. In this sense, we are implicitly showing that the unbroken 
NC generators interpolate states that are \emph{gapped}, %come in pairs and that what is really `broken' is some suitably defined time dependent generator for each one of these pairs. The interpolated state $|n,\vec{p} \, \rangle$ is now gapped 
in the sense that $E_n(0) \neq 0$. 

%In the following we further simplify the notation by dropping the sum over $n$, labeling the different sectors and we simply pose $|\pi(\vec p)\ket = |n, \, \vec p \rangle$, $E = E_n(0)$.

\vspace{.3cm}

%\section{C generators: gapless modes}
\noindent
{\em C-Generators: gapless modes.}
Eq.~\eqref{step3} should be compared with eq.~(6) of Nielsen and Chadha's paper~\cite{NC}. They find that the quantity that is constant in time is, in our notation,
\be \label{step4}
\kappa_{aI} = \sum_n  e^{- i E_n(0) t} \langle \mu| J^0_a   |n,0 \rangle \langle n,0  |  A_I |\mu\rangle - {\rm c.c.} 
\ee
The implicit assumption leading to~\eqref{step4} is that all currents evolve in time with the non-relativistic Hamiltonian  of which $|\mu \ket$ is the ground state ($\tilde H$). This assumption is violated by our system. However, our eq.~\eqref{step3} does reduce to \eqref{step4} whenever $Q_a$ commutes with $Q$, since in this case
%\begin{equation}
$e^{i \mu Q t} J^0_a e^{-i \mu Q t}  =  J^0_a$.%
%\end{equation}
\footnote{Once the algebra for the charges is given, the charge-current commutators are uniquely determined up to possible contact terms that vanish at zero momentum, i.e., total spacial derivatives. Since we took the $\vec p \to 0$ limit, such possible extra terms will not matter for us. \label{footnote algebra}}

Note that for $|n,0\rangle = |\mu\rangle$ the combination \eqref{step4} just gives zero, because $J^0_a$ and $A_I$ are hermitian operators. In order for $\kappa_{aI}$ to be time-independent \emph{and} different from zero, there must exist a Goldstone state $| \pi, \vec p \, \ket$ other than $|\mu\rangle$, whose energy goes to zero in the zero-momentum limit, $E_\pi(0) = 0$. Moreover, the matrix elements  $\langle \mu | J^0_a \, |\pi, \vec p\, \ket$ and $\bra \pi, \vec p \, | \, A_I |\mu \rangle$ should be nonzero for such a  state: both the broken current and the order parameter $A_I$ have to interpolate the Goldstone excitation.

Beyond this basic argument, the detailed analysis of the number and nature of gapless Goldstone bosons follows closely that of~\cite{NC} and features all the subtleties considered therein (see also~\cite{brauner,brauner2,kaon} for more recent refinements). We have nothing to add to the existing analyses, other than emphasize that they apply here  to broken generators of the {\em C-type} only. Notice that among the C-generators we have $Q$ itself. We devoted the bulk of \cite{NP} to studying the physical properties of the associated Goldstone boson.
%For these, the conclusion is that the number of broken generators is equal to the number of gapless Goldstones with odd dispersion relations (such as the relativistic $E(p)\sim p$) plus \emph{twice} the number of those with even dispersion relations (such as the non-relativistic $E(p)\sim p^2$). 

%Finally, it is worth commenting briefly on the role played by $\tilde H$ inside the charged sector. The point is that the sum over intermediate momentum eigenstates~\eqref{completeness} hits $\tilde H$, and not $H$, at the exponent. Therefore, it is natural to define the evolution operator within the charge sector by using $\tilde H$ rather than $H$. For instance, by means of the interpolating current we can define (see~\cite{NP} for more detail) a field operator $\pi(x)$,
%\begin{equation}
%\langle c|\pi(t,{\vec x})|\pi({\vec p})\rangle \sim e^{i({\vec p}\cdot {\vec x} - E(p) t)}\, .
%\end{equation}
%Since here $\tilde H |\pi({\vec p})\rangle = (E(p) + \lambda)|\pi({\vec p})\rangle $, it is implicit in the equation above that $\pi$ evolves with $\tilde H$ rather than with $H$, i.e. $\pi(t, {\vec x}) = e^{i \tilde H t} \pi(0, {\vec x})  e^{-i \tilde H t}$.
%
%

\vspace{.3cm}

%\section{NC generators: the gap}
\noindent 
{\em NC-Generators: the gap.}
Let us now consider the case where $Q_a$ does not commute with $Q=Q_1$. For our purposes, it is useful to write the commutation relations in hybrid form with charges and currents\footref{footnote algebra}:
%\footnote{Since the analog has to hold at the level of the charges, $[Q_a,Q_b] = i f^c_{ab} Q_c$,
%the only possible additions to \eqref{structure} are terms that vanish once we integrate in $d^3 x$---that is, total spacial derivatives. However, since we did integrate in $d^3 x$, or equivalently, we took the $\vec p \to 0$ limit, such possible extra terms will not matter for us either.\label{footnote_algebra}}:
\be \label{structure}
[Q_a, J^0_b(x)] = i f_{ab}^c \, J_c^0 (x) \; ,
\ee
where $f_{ab}^c$ are the group's structure constants, which are real for any Lie group.
We can now go back to eq.~\eqref{step3}. After expanding the exponentials on both sides of the current, using recursively \eqref{structure} to eliminate all $Q$'s, and re-exponentiating the result, we find  
\begin{equation}
e^{i \mu\, Q t} J^0_a e^{-i \mu\, Q t}  = (e^{-f_1 \, \mu t} )^b_a J^0_b \; , \label{adjoint}
\end{equation}
where $f_1$ is a matrix with entries $f_{1a}^b$.
That is, $i f_1$ is the adjoint representation of $Q_1$,
%\be
$(Q_1^A)^b_a = i f_{1a}^b$.
%\ee 
Eq.~\eqref{adjoint} above is nothing but the usual statement---applied to the currents---that the generators of a group live in the adjoint representation of that group.

The exponential acting on the current is now a finite dimensional matrix that `mixes' in a time-dependent fashion  the different currents of the group:
\be \label{exponential}
\kappa_{aI} = \sum_n e^{- i  E_n (0) t}  \, (e^{-f_1 \, \mu t} )^b_a \, \langle \mu | J^0_b   |n,0 \rangle \langle n,0|  A_I |\mu \rangle - {\rm c.c.} 
\ee
%For simplicity, we are dropping the sum over $n$---we are assuming that there is only one zero-momentum state that contributes to $\kappa_{aI}$, which we call $| \pi(0) \ket$.
We will now assume---as usual---that the symmetry group under consideration is the direct product of simple compact Lie groups ($SU(n)$, $SO(n)$, etc.), and of $U(1)$ factors. In this case the structure constants $f_{ab}^c$ can be taken to be totally antisymmetric---see e.g.~\cite{weinberg}.
Since $f_{1a}^b$ is real and antisymmetric, its eigenvalues are either zero or pure imaginary. The imaginary eigenvalues come in pairs $(+i q_a, -i q_a)$, with corresponding hermitian-conjugate pairs of  eigenvectors, defining a (non-hermitian) basis of generators in which
\be \label{diag Q_1}
f_{1a}^b = i \cdot {\rm diag}(0, \dots, 0,  q_1 , - q_1,  q_2, -q_2, \dots) \, .
\ee
By acting separately on each $\pm q_a$ block , it is straightforward to define instead an \emph{hermitian} basis, in which $ f_{1a}^b $ is real and block-diagonal, with $2\times2$ blocks  of the form
\be \label{block}
\left( \begin{array}{cc}
0 &  + q_a \\
- q_a & 0 \end{array}\right)\, .
\ee 
Let's assume that we started with eq.~\eqref{step2} directly in this hermitian basis where  $ f_{1a}^b $ is block-diagonal, and let's restrict our analysis to the spontaneously broken generators, which yield non-vanishing $\kappa_{aI}$.
If $J^0_a$ corresponds to a vanishing eigenvalue of $Q_1^A = i f_1$---i.e., if it commutes with $Q_1$---then we go back to the C-generator case and we conclude that $J^0_a$ interpolates a gapless particles.
If on the other hand $J^0_a$ is either of the paired currents acted upon by a $2\times2$ block of the form \eqref{block}, then
the exponential in \eqref{exponential} mixes its matrix elements with its companion's, with frequency $\mu q_a$.
%On the other hand, for $q_\alpha \neq 0$, 
%the exponential in~\eqref{exponential} mixes in a time dependent linear combination matrix elements of \emph{pairs} of currents, those acted upon by a $2\times2$ block of the form~\eqref{block}. 
In this case, the time-independence of $\kappa_{aI}$ implies a non-zero value for $E_n(0)$. This is most easily seen by using a complex notation:
$J^0_a$ can be expressed as $J^0_a = \hat J^0_a + \hat J^0_a {}^\dagger$---where $\hat J^0_a$ and $\hat J^0_a {}^\dagger$ are an hermitian-conjugate pair of non-hermitian generators that diagonalize $f_1$, as in \eqref{diag Q_1}---and eq.~\eqref{exponential} simply becomes
\begin{align}
\kappa_{aI} & =  \sum_n e^{- i  (E_n(0)  - \mu q_a )t} \, \langle \mu| \hat J^0_a   |n,0\rangle \langle n,0|  A_I |\mu \rangle  \label{final kappa}\\
& + e^{- i  (E_n(0)  + \mu q_a )t} \, \langle \mu | \hat J^0_a {}^\dagger   |n,0 \rangle \langle n,0|  A_I | \mu \rangle - {\rm c.c.} \nonumber 
\end{align}
Notice that we are not implicitly summing over $a$. Without loss of generality, we will assume that $\mu \, q_a$ is positive and $- \mu \, q_a$ negative.

By assumption, our  $| \mu \ket$ state is the ground state of the unbroken $\tilde H$ Hamiltonian. Therefore, there cannot be states with negative $E_n(\vec p)$. So, for the combination \eqref{final kappa} to be constant in time {\em and} different from zero, two conditions have to be met:
%\begin{enumerate}
%\item
{\em (i)} There exists a  state $| \pi_a, \vec p \, \ket$ whose energy in the zero momentum limit is
\be \label{gap}
E_a (0)= \mu \, q_a  \; .
\ee
This is our main result: the gap for the Goldstone excitations associated with the broken NC-generators.
%\item
{\em (ii)} Both $\hat J^0_a $ and $A_I$ are interpolating fields for such a state, in the sense that 
%(reinstating the operators' arguments)
%\be
$\langle \mu | \hat J^0_a |\pi_a, \vec p \, \rangle \neq 0$ and  $\langle \mu | A_I |\pi_a, \vec p \, \rangle \neq 0 $
%\ee
(recall that $A_I$ is hermitian.)
Notice that we have {\em one} gapped Goldstone mode for each {\em pair} of broken NC-generators. This is reminiscent of the Nielsen-Chadha counting~\cite{NC,brauner}. In fact, we still find that for even dispersion relations the number of broken generators is twice the number of Goldstones, just in the broader sense that gapped states have an `even' dispersion relation, i.e. $E(p) \sim |\vec p\, |^0 + {\cal O}(p^2)$. 
An intuitive picture of what is going on is provided by the following example.

\vspace{.3cm}

%\section{The SO(3) example}
\noindent
{\em The linear $SO(3)$ sigma model.}
Consider a Lagrangian with internal symmetry $SO(3)$, linearly realized on a scalar triplet $\vec \phi$:
\begin{equation} \label{lagrangian}
{\cal L} = -\sfrac{1}{2} \partial_\mu \vec \phi \cdot \partial^\mu \vec \phi -\sfrac{1}{2} m^2 |{\vec \phi} \, |^{2} - \sfrac{1}{4} \lambda |{\vec \phi} \, |^4 \, .
\end{equation} 
Let's pick one of the generators of $SO(3)$, say $\tau = \tau_3$, where $\tau_i$ generates rotations about the $\phi_i$ axis,
%\begin{equation}\nonumber
%\tau_1 = 
%\begin{pmatrix} 
%0 & 0 & 0\\
%0& 0 & 1\\
%0&-1&0
%\end{pmatrix} , \quad
%\tau_2 = 
%\begin{pmatrix} 
%0 & 0 & 1\\
%0& 0 & 0\\
%-1&0&0
%\end{pmatrix} , \quad
%\tau_3 = 
%\begin{pmatrix} 
%0 & 1 & 0\\
%-1& 0 & 0\\
%0&0&0
%\end{pmatrix} ,
%\end{equation}
and let's consider the theory at finite charge density for the corresponding charge $Q$. The standard way to build the non-relativistic Lagrangian at finite density---see e.g.~\cite{lebellac,son}---is  to introduce a constant non-dynamical gauge field pointing in the time direction, which in our case amounts to the
% Since $Q$ acts as a $U(1)$ subgroup on the components $\phi_1$ and $\phi_2$ only, we can switch to a complex parameterization $\Phi = \phi_1+i \phi_2$ for the $\phi_1, \phi_2$ plane, and introduce a chemical potential via the 
replacement
\begin{equation}
- \partial_\mu \Phi^\dagger \partial^\mu \Phi \rightarrow (\partial_0 - i \mu)  \Phi^\dagger (\partial_0 + i\mu) \Phi - \partial_j  \Phi^\dagger \partial_j  \Phi\, ,
\end{equation}
where $\Phi$ is the complex combination $\Phi = \phi_1+i \phi_2$.
After going to polar coordinates, $\Phi = \sigma e^{i \theta}$, we find
\begin{align}
{\cal L}_{(\mu)}=& -\sfrac{1}{2} \partial_\mu \sigma \partial^\mu \sigma -\sfrac{1}{2} \sigma^2 \partial_\mu \theta \partial^\mu \theta - \sfrac{1}{2}  \partial_\mu \phi_3 \partial^\mu \phi_3 \label{lag2} \\ 
& - \sfrac{1}{2} m^2 \, (\sigma^2 +\phi_3^2) - \sfrac1{4}{\lambda} \, (\sigma^2 +\phi_3^2)^2 + \mu\,  \sigma^2 \dot \theta + \sfrac{1}{2} \mu^2 \sigma^2
\nonumber 
\end{align}
The ground state at finite chemical potential corresponds to constant field solutions of this Lagrangian. While we always have $\langle \phi_3 \rangle = 0$, there are cases when $\langle \sigma \rangle \neq 0$, which spontaneously breaks the symmetry generated by $Q$. This happens when the configuration $\vec\phi=0$ was unstable to begin with ($m^2<0$), or when the chemical potential exceeds a critical value ($\mu^2 > m^2$).

In the broken phase, the VEV of the radial field is 
%\begin{equation}
$\langle\sigma\rangle ^2 = \frac1\lambda (\mu^2 - m^2)$.
%\end{equation}
Expanded to second order about this configuration, the Lagrangian~\eqref{lag2} reads 
\begin{align} \nonumber
{\cal L}_{(\mu)} &\simeq  -\sfrac{1}{2} \partial_\mu \delta \sigma \partial^\mu \delta \sigma -\sfrac{1}{2} \langle\sigma\rangle^2 \partial_\mu \theta \partial^\mu \theta - \sfrac{1}{2}  \partial_\mu \phi_3 \partial^\mu \phi_3 \\ 
& + 2 \mu  \langle\sigma\rangle \dot \theta \delta \sigma -(\mu^2 - m^2) \delta \sigma^2 - \sfrac{1}{2} \mu^2 \phi_3^2  \label{L quad}\, .
\end{align}

Let's assume that $\langle \vec \phi \rangle$ points in the $\phi_1$ direction.
Of the whole $SO(3)$ symmetry group, the only residual symmetry is that associated with $\tau_1$---rotations about $\phi_1$.
This symmetry breaking pattern would normally be associated with two massless Goldstone bosons, one for each broken generator. Instead, here we see that the would-be Goldstone field $\phi_3$ associated with  $\tau_2$ has acquired a mass $m_3= \mu$, in agreement with our general result, since $\tau_2$ does not commute with $\tau_3$. The angular field ($\theta$) {\em is} massless, and can be identified with the Goldstone boson associated with $\tau_3$, which obviously commutes with itself.

In the alternative (but equivalent) SSP language \cite{NP}, one starts from the relativistic Lagrangian~\eqref{lagrangian}, and looks for a {\em time-dependent} background solution that rotates about the $\phi_3$ axis, with
%\be
$\dot{\vec \phi} = \mu \, \tau_3 \cdot \vec \phi$.
%\ee
This just corresponds to a constant speed in the angular field, $\theta(x) = \mu \, t$. Such solutions are allowed only for $m^2<0$ or for $\mu^2 > m^2$, in agreement with what we found above. After expanding in $\theta$ fluctuations  about such a solution, $\theta \to \mu \, t + \theta$, one finds precisely the Lagrangian~\eqref{lag2}, and its quadratic approximation \eqref{L quad}, with the same excitation spectrum as above.
%However, by writing the time-dependent SSP background in terms of the original fields, 
%\be
%\phi_1 = \langle\sigma\rangle \sin \mu t, \qquad \phi_2 = \langle\sigma\rangle \cos \mu t, \qquad \phi_3 =0 \; ,
%\ee
%it is not difficult to see what is happening in terms of the generators. 
%In this language, all $SO(3)$ generators are broken. $\tau_3$ interpolates a massless state. This is the excitation in the direction of motion in field space or, equivalently, that related to the field $\theta$. As to $\tau_1$ and $\tau_2$, they are broken in a time dependent fashion. we can build a time-dependent combination of them that `follows' the SSP solution and generates, instantaneously, rotations about the corresponding axis. This is unbroken by $\langle \vec \phi \rangle$, but broken by $\langle \dot{\vec \phi} \rangle$. The orthogonal time-dependent combination---that maximally broken by $\langle \vec \phi \rangle$---moves the system in the $\phi_3$ direction, and interpolates the associated massive Goldstone.
However, the SSP language is closer to the viewpoint we emphasized in this paper, because it makes manifest that having a finite charge density for a spontaneously broken charge, necessarily implies a spontaneous breakdown of time-translations as well. Moreover, it stresses that Lorentz-invariance is broken spontaneously, by the field configuration one considers, rather than at the level of the Lagrangian. And finally, it gets the breaking pattern for internal symmetries right: {\em all} $SO(3)$ generators are broken, albeit in a time-dependent fashion, in the sense that $\langle [\tau_1, \vec \phi(x)] \rangle$ and $\langle [\tau_2, \vec \phi(x)] \rangle$ depend on time. This is what one would discover from our general analysis above, if one evaluated the order parameters $A_I$ at generic positions $x$ rather than at the origin.

\vspace{.3cm}

%\section{Concluding remarks}
\noindent
{\em Concluding remarks.}
We conclude by stressing that our derivation of eq.~\eqref{gap} involved no approximation. As a result, our expression for the gap is an exact non-perturbative prediction. 
Our results evade the Nielsen-Chadha theorem, because one of its (implicit) assumptions is violated, namely that  $Q$ commute with all other broken charges. This crucial difference should be taken into consideration also when comparing our results with the literature, like for instance the ``kaon condensation" model of refs.~\cite{kaon2,kaon}.
On the other hand, our results agree with the theorem of \cite{kaon}, which states---among other things---that  there are no subtleties in counting the gapless Golstones for relativistic theories as long as all charge densities vanish.

\vspace{.3cm}

\noindent
{\em Acknowledgements.}
The work of AN is supported by the DOE (DE-FG02-92-ER40699) and by NASA ATP
(09-ATP09-0049).

%%%%%%%%%%%%%%%%%%%%%%%%%%%%%%%
%%%%%%%%%%%%%%%%%%%%%%%%%%%%%%%

\end{document}